%% file: main.tex
\pdfoutput=1
\documentclass[manuscript,pdf]{acmart}

\AtBeginDocument{%
  \providecommand\BibTeX{{%
    \normalfont B\kern-0.5em{\scshape i\kern-0.25em b}\kern-0.8em\TeX}}}

\setcopyright{rightsretained}
\copyrightyear{2024}

\acmConference[CARS Workshop @ RecSys]{2024}{Italy}

\usepackage[titletoc]{appendix}
\usepackage{cleveref}
\usepackage{amsmath}
\usepackage{paralist}
\usepackage{subcaption}
\usepackage{multirow}
\usepackage{tikz}
\usepackage[inline]{enumitem}
\usepackage{pgfplots}

\input{math_commands}

\begin{document}

\title{\textsc{SimRec}: Mitigating the Cold-Start Problem in Sequential Recommendation by Integrating Item Similarity}

\author{Shaked Brody}
\affiliation{%
  \institution{Amazon}
}
\email{shakedbr@amazon.com}
\orcid{0009-0004-4839-4341}

\author{Shoval Lagziel}
\affiliation{%
  \institution{Amazon}
}
\email{shovall@amazon.com}
\orcid{0000-0002-1657-2076}


\newcommand{\slfrac}[2]{\left.#1\middle/#2\right.}
\newcommand{\simrec}{\textsc{SimRec}}

\input{00-abstract}

\maketitle

\input{01-intro}

\input{03-cold-start}

\input{04-method}
\input{05-experiments}

\input{06-concolusion}
\section*{Acknowledgments}
We would like to thank our team and in particular Iftah Gamzu and Yonathan Aflalo for the helpful discussions.
\clearpage
\bibliographystyle{abbrvnat}
\bibliography{references}

\clearpage

\appendix

\input{appendix}

\end{document}

%% file: math_commands.tex
\usepackage{amsmath,amsfonts,bm}

\def\eqref#1{equation~\ref{#1}}

\def\1{\bm{1}}

\def\vf{{\bm{f}}}

\def\vr{{\bm{r}}}
\def\vs{{\bm{s}}}

\def\vv{{\bm{v}}}

\def\v1{{\vec{\bm{\mathbbm{1}}}}}

\def\mE{{\bm{E}}}

\DeclareMathAlphabet{\mathsfit}{\encodingdefault}{\sfdefault}{m}{sl}
\SetMathAlphabet{\mathsfit}{bold}{\encodingdefault}{\sfdefault}{bx}{n}

%% file: 00-abstract.tex
\begin{abstract}
    Sequential recommendation systems often struggle to make predictions or take action when dealing with cold-start items that have limited amount of interactions.
    In this work, we propose \simrec{} -- a new approach to mitigate the cold-start problem in sequential recommendation systems.    
    \simrec{} addresses this challenge by leveraging the inherent similarity among items, incorporating item similarities into the training process through a customized loss function. Importantly, this enhancement is attained with identical model architecture and the same amount of trainable parameters, resulting in the same inference time and requiring minimal additional effort. 
    This novel approach results in a robust contextual sequential recommendation model capable of effectively handling rare items, including those that were not explicitly seen during training, thereby enhancing overall recommendation performance. 
    Rigorous evaluations against multiple baselines on diverse datasets showcase \simrec{}'s superiority, particularly in scenarios involving items occurring less than 10 times in the training data. The experiments reveal an impressive improvement, with \simrec{} achieving up to 78\% higher HR@10 compared to SASRec.
     Notably, \simrec{} outperforms strong baselines on sparse datasets while delivering on-par performance on dense datasets.   
     Our code is available at \url{https://github.com/amazon-science/sequential-recommendation-using-similarity}.

\end{abstract}

%% file: 01-intro.tex
\section{Introduction}\label{Se:introduction}

Sequential recommendation systems play a vital role in diverse applications, including e-commerce websites and video streaming platforms, providing valuable recommendations that are tailored to the user's preferences.
These systems model each user by past interactions and use this information to predict the next item.

Many efforts have been made to improve modeling methods in sequential recommendations. 
\citet{hidasi2015session} used recurrent neural networks (RNNs) to process user history sequentially.
Later, SASRec \cite{kang2018self} used the self-attention mechanism of the Transformer architecture \cite{vaswani2017attention} to predict the next item.
While some approaches make predictions based solely on the sequence of item IDs (i.e. \textit{ID-based methods} \cite{yuan2023go}), others take advantage of additional contextual features.
TiSASRec \cite{li2020time} considered time intervals between actions in the user's history when predicting the next recommendation.
CARCA \cite{rashed2022context} used additional attributes such as the item's brand and other dense features. %
Furthermore, several attempts incorporated self-supervised learning techniques into their training process.
BERT4Rec \cite{sun2019bert4rec} adopted BERT's \cite{devlin2018bert} pre-training objective in the sequential recommendation setting.
Similarly, as done in CARCA, S3-Rec \cite{zhou2020s3} employed item attributes, but in a self-supervised pre-training phase.

In recommendation systems, the cold-start problem, characterized by reduced performance when dealing with items with sparse or no appearances, is a longstanding challenge, particularly in the context of sequential recommendation \cite{8229786,10061072}.
\input{figures/beauty_freq_hist}

\Cref{Fig:beauty_freq_hist} shows an example of the cold-start problem in the Amazon Beauty dataset \cite{he2016ups,mcauley2015image}, in which 91\% of the items appear less than 10 times in the training set. 
This means that most items are infrequent, limiting their contribution to the recommendation system's training and leading to poor performance in predicting them during test time.

To mitigate this limitation, we propose \simrec{}.
Our method leverages the fact that many distinct items are similar to each other.
In particular, rare items may share traits with abundant items that are frequently seen during training.
For example, consider the following items from the Amazon Beauty dataset, which differ only in size: ``NOW Foods Organic Lavender Oil, \textbf{1} ounce'' and ``NOW Foods Organic Lavender Oil, \textbf{4} ounce''.
\simrec{} exploits this observation by incorporating a \emph{similarity distribution} for each item, describing how similar the item is to other items in the dataset.
In this work, we employ a text embedding model \cite{li2023towards} to represent the items and subsequently calculate the similarity between them. It is worth noting that there are no technical limitations regarding the choice of similarity metric, offering flexibility for different scenarios and datasets.
We introduce a new loss function that incorporates this similarity distribution during training, allowing the model to better learn on cold-start items even if they are not explicitly seen during training\footnote{By saying ``not explicitly seen'' we mean that the item has no interactions with users in the training data, but is included in the item similarity data.
}.
Our approach is simple, requires no additional learnable parameters, and as a result, maintains the same inference time and complexity while significantly enhancing recommendation performance across a wide range of datasets.

Recent studies have tackled the cold-start problem from different perspectives. 
The Recformer model \cite{RecformerMcAuley} adopts the Reformer architecture \cite{reformer}, designed to handle long sequences, for recommendation purposes.
Positioned as a \textit{Text-Only Method}, Recformer processes sentence inputs for each item, leveraging language representations. 
Although it has demonstrated superior performance, it comes with a trade-off in computational complexity compared to \textit{ID-based methods}.
\simrec{} inference time is identical to an \textit{ID-based method} (i.e., SASRec\cite{kang2018self}). Its computational cost is evident in the similarity calculation phase which happens once before training and in the training phase during the computation of our new loss.
Several works have explored the use of meta-learning frameworks to address the cold-start problem \cite{zheng2021cold,huang2022learning}. 
However, this line of works differ from ours as they focus on scenarios where a history of interactions and a small set of cold-start items are provided during inference, enabling the system to recommend one item from this limited set.
In contrast, we work in a distinct setting, where given a user's history, the objective is to recommend an item from the entire item pool. 
This approach allows us to predict both rare and abundant items, offering a comprehensive solution to the cold-start problem in recommendation systems.
A few works tackled the cold-start problem within domain specific datasets \cite{chou2016addressing,deldjoo2019movie,ren2019financial,bi2020dcdir}. 
Our approach is applicable to a wider array of domains. 
It only requires item similarity data, which can be calculated based on basic features such as item title that are commonly available. 
This eliminates the need for labor-intensive feature curation specific to each dataset.

%% file: figures/beauty_freq_hist.tex
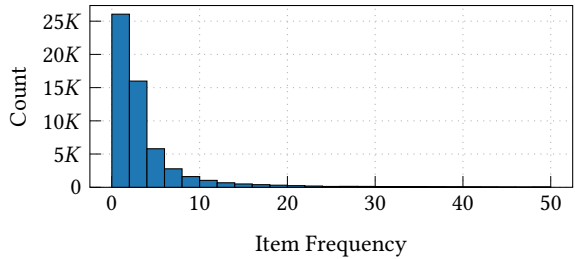
\begin{figure}
    \centering
\begin{tikzpicture}

\definecolor{darkgray176}{RGB}{176,176,176}
\definecolor{steelblue31119180}{RGB}{31,119,180}

\begin{axis}[
height=4cm,
width=8cm,
tick pos=left,
x grid style={darkgray176},
grid style={dotted, gray},
tick align=inside,
xmajorgrids,
xlabel={Item Frequency},
xmin=-2.5, xmax=52.5,
xtick style={color=black},
y grid style={darkgray176},
ylabel={Count},
ymin=0, ymax=27354.6,
ytick style={color=black},
ymajorgrids,
scaled ticks=false,
tick label style={/pgf/number format/fixed},
ytick={0,5009,10000,15000,20000,25000},
ylabel near ticks,
yticklabels={
  \(\displaystyle {0}\),
  \(\displaystyle {5K}\),
  \(\displaystyle {10K}\),
  \(\displaystyle {15K}\),
  \(\displaystyle {20K}\),
  \(\displaystyle {25K}\)
}
]
\draw[draw=black,fill=steelblue31119180,fill opacity=1] (axis cs:0,0) rectangle (axis cs:2,26052);
\draw[draw=black,fill=steelblue31119180,fill opacity=1] (axis cs:2,0) rectangle (axis cs:4,15983);
\draw[draw=black,fill=steelblue31119180,fill opacity=1] (axis cs:4,0) rectangle (axis cs:6,5802);
\draw[draw=black,fill=steelblue31119180,fill opacity=1] (axis cs:6,0) rectangle (axis cs:8,2783);
\draw[draw=black,fill=steelblue31119180,fill opacity=1] (axis cs:8,0) rectangle (axis cs:10,1614);
\draw[draw=black,fill=steelblue31119180,fill opacity=1] (axis cs:10,0) rectangle (axis cs:12,1045);
\draw[draw=black,fill=steelblue31119180,fill opacity=1] (axis cs:12,0) rectangle (axis cs:14,682);
\draw[draw=black,fill=steelblue31119180,fill opacity=1] (axis cs:14,0) rectangle (axis cs:16,503);
\draw[draw=black,fill=steelblue31119180,fill opacity=1] (axis cs:16,0) rectangle (axis cs:18,397);
\draw[draw=black,fill=steelblue31119180,fill opacity=1] (axis cs:18,0) rectangle (axis cs:20,317);
\draw[draw=black,fill=steelblue31119180,fill opacity=1] (axis cs:20,0) rectangle (axis cs:22,250);
\draw[draw=black,fill=steelblue31119180,fill opacity=1] (axis cs:22,0) rectangle (axis cs:24,191);
\draw[draw=black,fill=steelblue31119180,fill opacity=1] (axis cs:24,0) rectangle (axis cs:26,129);
\draw[draw=black,fill=steelblue31119180,fill opacity=1] (axis cs:26,0) rectangle (axis cs:28,144);
\draw[draw=black,fill=steelblue31119180,fill opacity=1] (axis cs:28,0) rectangle (axis cs:30,133);
\draw[draw=black,fill=steelblue31119180,fill opacity=1] (axis cs:30,0) rectangle (axis cs:32,98);
\draw[draw=black,fill=steelblue31119180,fill opacity=1] (axis cs:32,0) rectangle (axis cs:34,87);
\draw[draw=black,fill=steelblue31119180,fill opacity=1] (axis cs:34,0) rectangle (axis cs:36,81);
\draw[draw=black,fill=steelblue31119180,fill opacity=1] (axis cs:36,0) rectangle (axis cs:38,74);
\draw[draw=black,fill=steelblue31119180,fill opacity=1] (axis cs:38,0) rectangle (axis cs:40,67);
\draw[draw=black,fill=steelblue31119180,fill opacity=1] (axis cs:40,0) rectangle (axis cs:42,62);
\draw[draw=black,fill=steelblue31119180,fill opacity=1] (axis cs:42,0) rectangle (axis cs:44,55);
\draw[draw=black,fill=steelblue31119180,fill opacity=1] (axis cs:44,0) rectangle (axis cs:46,44);
\draw[draw=black,fill=steelblue31119180,fill opacity=1] (axis cs:46,0) rectangle (axis cs:48,37);
\draw[draw=black,fill=steelblue31119180,fill opacity=1] (axis cs:48,0) rectangle (axis cs:50,48);
\end{axis}

\end{tikzpicture}
\vspace{-0.2cm}
    \caption{The Cold-Start Problem: The item frequency of the Amazon Beauty training set. 91\% of the items appear less than 10 times. }
    \label{Fig:beauty_freq_hist}
\vspace{-0.5cm}
\end{figure}

%% file: 03-cold-start.tex
\section{Cold-Start Items and Density}\label{Se:cold-start}

Cold-start items are items that have few interactions in the dataset or are entirely absent, meaning they have been rarely interacted by users.
We define the \textit{density} of a dataset as the average number of interactions per item:
\begin{equation}
    Density(\mathcal{D}) = \frac{\sum_{u\in \mathcal{U}}|u|}{|\mathcal{V}|} = \frac{\# interactions}{\# items}
\end{equation}

Where $\mathcal{U}$ is the set of users, $|u|$ is the number of interactions per user, and $\mathcal{V}$ is the set of items.
We can say that the sparsity of a dataset increases when its density decreases, i.e., a dataset is said to be sparse when its density value is relatively small.

An important observation is that although sparse datasets contain many rare items, many of them are similar.
\input{figures/similar_items_example_beauty}
\Cref{Fig:similarity-example-beauty} shows an example of an item from the Amazon Beauty dataset and its most similar items. Note that while the item is relatively frequent, it is similar to rare items.
This observation can be used to improve performance on rare items, ultimately leading to overall performance improvement, especially on sparse datasets.
In \Cref{Se:method}, we explain how we measure the similarities between items and how we incorporate this information into the training process.

%% file: figures/similar_items_example_beauty.tex
\begin{table*}[ht!]
\centering
\caption{
Example of an item and its most similar items from the Beauty dataset. 
}
\setlength{\tabcolsep}{5pt}
\begin{tabular}{llc}
\toprule
& Item   & Frequency \\  
\midrule
 & \textbf{Olay Pro-X Advanced Cleansing System 0.68 Fl Oz, 1-Count} & \textbf{351}\\
k=1 & Olay Pro-X Microdermabrasion Plus Advanced Cleansing System Refill, 1-Count, 1.7 fl oz & 8\\
k=2 & Olay Professional Pro-X Advanced Cleansing System Set & 12\\
k=3 & Olay Professional Pro-X Restorative Cream Cleanser, 5 Ounce & 1\\
k=4 & Olay Pro-X Microdermabrasion Plus Advanced Cleansing System, 1-Kit & 63\\
k=5 & Olay Age Defying Classic Cleanser , 6.8-Fluid Ounce  (Pack of 4) & 2\\
\midrule
\end{tabular}
\label{Fig:similarity-example-beauty}
\end{table*}

%% file: 04-method.tex
\section{Methodology}\label{Se:method}
In this section, we elaborate on \simrec{} -- our proposed approach.
We begin by describing the computation of item similarities.
Then, we outline how \simrec{} incorporates similarity distributions during the training process, utilizing a new loss function.

\vspace{-0.1cm}
\subsection{Item Similarity Distribution}\label{Se:item-similarity}
\simrec{} requires a similarity metric between pairs of items, which is then used to produce the similarity distribution for each item.
One way to find item similarities is by considering textual features, and calculating the similarity between their dense representation. 
Large Language Models (LLMs) have demonstrated promising results in many Natural Language Processing (NLP) tasks such as language modeling, question answering and sentiment analysis \cite{radford2019language, brown2020language, raffel2020exploring}.
The notion of generating dense text representations has recently gained significant attention, as these representations hold great potential for various tasks such as semantic search, text retrieval and text similarity \cite{reimers2019sentence,reimers2020curse}.

Specifically, our process involves applying a pre-trained text embedding model to the item titles, resulting in a dense representation for each item. Subsequently, we compute the similarity between two items by utilizing the cosine similarity between their respective representations.
Notably, while we employed text for similarity calculations, our method is not limited to this modality. Without inherent constraints, our suggested loss can be applied to any modality and any similarity function.

Initially, we generate the embedding for each item title, denoted as $\vv_i$. Then, we proceed to assess the similarities of the $i^{th}$ item with all other items in the dataset, resulting in a similarity vector $\vs^i \in \mathbb{R}^{|\mathcal{V}|}$.
Then, to create a valid distribution, we apply the softmax function to $\vs^i$ resulting in $\widehat{\vs^i}$ that describes how similar the $i^{th}$ item is to the other items.
We incorporate these distributions during the training process, as we explain next. %

\vspace{-0.1cm}
\subsection{New Loss Function}\label{SubSe:new-loss-fucntion}
Since our model architecture is equivalent to SASRec \cite{kang2018self}, we refer the reader to \citet{kang2018self} for more details.

Given a user $u \in §\mathcal{U}$ and the user-item interaction history sequence $h_u=\left[a_1^{u}, a_2^{u}, ..., a_{|u|}^{u}\right]$, the model outputs the sequence $\left[\vf_1, \vf_2, ... , \vf_{|u|}\right]\in\mathbb{R}^{d \times |u|}$ such that $d$ is the hidden size and $\vf_t$ corresponds to the t$^{th}$ item of the input sequence.
To determine the relevance score of an item $i\in \mathcal{V}$ as the next item after the $t^{th}$ item of the sequence $h_u$ (considering the initial $t$ items), we multiply $\vf_{t}$ with $\mE_i$ (the embedding of $i^{th}$ item that is been learned during training) which results in $r^t_i = \left<\vf_t, \mE_i \right>$.
A high score indicates a high level of relevance. Based on the ranking of these scores, we can provide recommendations.

\simrec{} uses a new loss function $\mathcal{L}_{\simrec{}}$ that combines
\begin{inparaenum}[(a)]
\item binary cross-entropy loss, and
\item \emph{similarity} loss. 
\end{inparaenum}
We first describe the binary cross-entropy loss which is commonly used in training sequential recommendation systems \cite{kang2018self,li2020time}, then our proposed \emph{similarity} loss, and finally $\mathcal{L}_{\simrec{}}$ that combines the two. 

We define $h_u^+=\left[a_2^{u}, ..., a_{{|u|}}^{u}\right]$ to be the positive sequence that represents the next true item for any prefix of $h_u$, as the left-shifted sequence $h_u$.
We also define $h_u^-=\left[a_1^{-u}, ..., a_{|u|-1}^{-u}\right],\forall i, a_i^{-u} \notin h_u$ to be the negative sequence that consists of items that are not part of $h_u$, which are selected randomly during training.
Now, we define the binary cross-entropy loss as follows:
\begin{equation}
    \mathcal{L}_{BCE} = -\sum_{u \in \mathcal{U}} \sum_{t=1}^{|u|-1} \left[log\left(\sigma\left(r^t_{ a_t^{+u}}\right)\right)+log\left(1-\sigma\left(r^t_{ a_t^{-u}}\right)\right)\right]
\end{equation}
where $\sigma$ is the sigmoid function.

To incorporate item similarity data, we define \emph{similarity} loss as the cross-entropy loss between the model output relevancy distribution over all items and the similarity distribution.

We denote $\vr^t=[r^t_{1},...,r^t_{|\mathcal{V}|}]$ as the model output relevancy scores for the next item after the $t^{th}$ element of $h_u$.
To maintain a distribution, we apply the softmax function to $\vr^t$ resulting in $\widehat{\vr^t}$.
We denote $\widehat{\vs^{a_t^{+u}}}$ as the similarity distribution of the next true item $a_t^{+u}$.

We define the similarity loss as follows:
\begin{equation}
    \mathcal{L}_{Similarity} = -\sum_{u \in \mathcal{U}} \sum_{t=1}^{|u|-1} \sum_{i=1}^{|\mathcal{V}|} \left[\widehat{\vs^{a_t^{+u}}_i}\cdot log\left(\widehat{r^t_i}\right)\right]
\end{equation}
The final loss function we optimize is:
\begin{equation}
    \mathcal{L}_{\simrec{}} = (1-\lambda) \cdot \mathcal{L}_{BCE} + \lambda \cdot \mathcal{L}_{Similarity}, 
    \,\,\,\,\text{~where~}\lambda \in [0,1]
\end{equation}

By using the similarity distribution during training, we actually incorporate new information into the model compared to other baselines such as SASRec \cite{kang2018self}.
In \Cref{Se:ablation}, we explore an alternative approach for integrating this information into the model through the item embedding, and demonstrate that the loss function we have developed yields superior results.

%% file: 05-experiments.tex
\section{Experiments}\label{Se:experiments}
We conduct a comprehensive analysis of our model's effectiveness, exploring:
\begin{inparaenum}[(a)]
\item the overall recommendation performance,
\item the cold-start performance, and
\item the correlation between the density of the datasets and the performance improvement facilitated by our approach.
\end{inparaenum}
Finally, we present an ablation study analyzing the different components of our approach. 

In all our experiments, we compute similarity between embeddings generated by the GTE-Large model \cite{li2023towards} applied to item titles. Further details on datasets and implementation can be found in \Cref{Se:datasets,Ap:hyper-parameter}.

\input{tables/results_table}

\subsection{Recommendation Performance}
Our evaluation covers both dense datasets (ML-1M, Steam \cite{harper2015movielens,kang2018self}) and sparse datasets (Tools, Beauty, Pet Supplies and Home \& Kitchen \cite{he2016ups,mcauley2015image}).
We compare \simrec{} performance to one na\"ive baseline (TopPop) and four sequential recommendation models. The latter include two \textit{ID-based methods} \cite{yuan2023go} (SASRec \cite{kang2018self} and BERT4Rec \cite{sun2019bert4rec}, representing items with IDs which are subsequently converted to embeddings), and two \textit{ID-based methods with additional context} (TiSASRec \cite{li2020time} and CARCA \cite{rashed2022context}), such as \simrec{}.

\Cref{Tab:results} shows the results of our experiments. 
\simrec{} outperforms on the sparse datasets -- Tools, Beauty, Pet Supplies and Home \& Kitchen, and is on-par with other strong-baselines \cite{kang2018self,li2020time,rashed2022context,sun2019bert4rec} on the dense datasets -- ML-1M and Steam.

CARCA \cite{rashed2022context}, which utilizes contextual features alongside item embeddings, provides competitive results on the sparse datasets, however, it achieves relatively lower performance on the dense datasets -- ML-1M and Steam. Nevertheless, \simrec{} excels on sparse datasets and achieves on-par performance as the other strong baselines on dense datasets.
This suggests that \simrec{} is compatible with a wider range of datasets compared to the other baselines, and CARCA in particular.

Since the architecture of \simrec{} is identical to that of SASRec \cite{kang2018self} and our approach differs in the optimization process, we show the relative improvement of \simrec{} compared to SASRec in \Cref{Tab:results}.
Notably, \simrec{} achieves up to \textbf{28.85\%} HR@10 improvement and up to \textbf{36.7\%} NDCG@10 improvement over SASRec.

\input{figures/cold_start_beauty_fig}
\input{figures/density_fig}

\subsection{Cold Start Performance}\label{Se:cold-start-experiments}
Our evaluation suggests that \simrec{}'s competitive performance arises from its capacity to enhance recommendations for cold-start items, which are more common in sparse datasets.
In this section, we compare the performance of \simrec{} and SASRec \cite{kang2018self}, both sharing the same architecture, specifically focusing on cold-start items.
\Cref{Fig:cold-start-beauty} shows the performance of \simrec{} and SASRec on the Beauty dataset \cite{he2016ups,mcauley2015image}.
Each point shows the HR@10 for samples in the test set where the true next item has a specific frequency (calculated on the training set).

First, it is worth noting that \simrec{} outperforms SASRec on cold-start items with low frequency.
The largest difference is evident when considering the rarest items that occur less than 10 times in the training set.
\simrec{} achieves \textbf{78\%} higher HR@10  and \textbf{101\%} higher NDCG@10 compared to SASRec for such items, accounting for 61\% of the test set (where \simrec{} achieves 0.405 HR@10 and 0.269 NDCG@10 compared to 0.227 and 0.134 achieved by SASRec, respectively).
For more abundant items, i.e., those occurring 10 times or more, the improvement is less significant, with a relative improvement of 8\% HR@10 and 16.6\% NDCG@10 (0.882 HR@10 and 0.673 NDCG@10 for \simrec{}, compared to 0.816 and 0.577, respectively, achieved by SASRec).
In addition, \simrec{} demonstrates the ability to predict items not explicitly seen during the training process (item frequency = 0), achieving a HR@10 of 0.253 and an NDCG@10 of 0.165, while SASRec fails completely with an HR@10 and an NDCG@10 of 0. 
Notable improvements with \simrec{} are evident for other sparse datasets and can also be found in \Cref{Se:cold-start-additional}.

\subsection{Density vs. Performance}\label{Se:density-experiments}

To investigate the relationship between the density of a dataset and the performance of \simrec{}, we generated several variants of the Beauty dataset \cite{he2016ups,mcauley2015image}, with different densities. %
Previous work \cite{kang2018self,rashed2022context,li2020time} has eliminated items occurring less than $n=5$ times in the raw dataset. 
We repeated this pre-processing method, and used different values of $n$  to derive dataset variants with different densities.
The statistics of the datasets can be found in \Cref{Se:density-addtional}.

\Cref{Fig:density} shows the relative performance gain of \simrec{} over SASRec \cite{kang2018self} on the different variants of the Beauty dataset.
We choose to present the relative gain since comparing the HR@10 and NDCG@10 between different datasets with different number of items has an effect on the evaluation process.
We find that the relative gain increases as the dataset becomes sparser.
Thus, the advantage of \simrec{} over SASRec decreases as the dataset becomes denser. For example, for the densest dataset ($n=20$), \simrec{} achieves an improvement of 18.3\% and 22.7\% for HR@10 and NDCG@10, respectively, while for the sparsest dataset ($n=0$), \simrec{} achieves an improvement of 31.8\% and 41.5\% for HR@10 and NDCG@10, respectively. %
These findings align with the results in \Cref{Tab:results}, where \simrec{} shows its most significant performance improvements on sparse datasets.

\vspace{-0.1cm}

\input{tables/ablation_table}

\vspace{-0.4cm}
\subsection{Ablation Study}\label{Se:ablation}
We examine the individual contributions of various components of our approach.
We conducted experiments using the following models on the Beauty dataset:
\begin{enumerate*}
    \item \textbf{SASRec} \cite{kang2018self},\label{2}
    \item \textbf{SASRec + embeddings}: the item embeddings are initialized with the embeddings generated by GTE-Large \cite{li2023towards},\label{3}
    \item \textbf{SASRec + embeddings + $\mathcal{L}_{\simrec{}}$}: employing our loss,\label{4} and
    \item \textbf{\simrec{}}.\label{5}
\end{enumerate*}
\Cref{Tab:ablation} summarizes the results.
Initializing SASRec embeddings with the GTE-Large embeddings (\ref{3}) results in a remarkable performance improvement of over 20\% HR@10 and 23\% NDCG@10 compared to SASRec (\ref{2}) that uses the Glorot initialization method \cite{glorot2010understanding} for the embedding weights.
Using the embeddings in this way is similar to CARCA \cite{rashed2022context} in the sense that additional data is explicitly introduced into the model.

Using our proposed method (\ref{5}) yields superior performance compared to embedding initialization (\ref{3}).
This approach (\ref{5}) results in a relative gain of over 28\% in HR@10 and 36\% in NDCG@10 compared to SASRec (\ref{2}).
Using both embedding initialization and our new loss (\ref{4}) leads to similar performance as using the new loss alone (\ref{5}).
This indicates that our new loss is the main contributor to the performance gain. 
Results regarding the choice of a text embedding model for item similarity are detailed in \Cref{Se:text-embedding-method}.

%% file: tables/results_table.tex
\begin{table}[!ht]
\centering
\caption{\simrec{} outperforms other strong baselines on sparse datasets, and achieves on-par performance on dense datasets. 
(*) marks statistical significance improvement compared to the second best model, with $p<0.01$.}
\setlength{\tabcolsep}{5pt}
\begin{tabular}{llcccccccc}
\toprule
\multirow{2}{*}{Dataset}  & \multirow{2}{*}{Metric}  & \multirow{2}{*}{TopPop} & \multirow{2}{*}{SASRec} & \multirow{2}{*}{TiSASRec} & \multirow{2}{*}{BERT4Rec} & \multirow{2}{*}{CARCA} & \multirow{2}{*}{\simrec{}} & \simrec{} & \multirow{2}{*}{Density}\\
& & & & & & & & vs. SASRec & \\ 

\midrule
\multirow{2}{*}{Tools} & HR@10     & 0.444 & 0.407 & 0.409 & 0.410 & \underline{0.501} & \textbf{0.517$^*$} & 27.03\% & \multirow{2}{*}{\hphantom{00}6.2}\\
                       & NDCG@10  & 0.261 & 0.252 & 0.252 & 0.242 & \underline{0.311} & \textbf{0.333$^*$} & 32.14\% &\\
\midrule                        
\multirow{2}{*}{Beauty} & HR@10   & 0.460 & 0.461 & 0.437 & 0.455 & \underline{0.559} & \textbf{0.594$^*$} & 28.85\%& \multirow{2}{*}{\hphantom{00}6.9} \\
                        & NDCG@10  & 0.263 & 0.311 & 0.289 & 0.294 & \underline{0.374} & \textbf{0.425$^*$} & 36.70\%& \\
\midrule
Pet        &  HR@10    & 0.483 & 0.542 & 0.536 & 0.536 & \underline{0.618} & \textbf{0.632$^*$} & 16.61\% & \multirow{2}{*}{\hphantom{00}9.1} \\
Supplies   & NDCG@10  & 0.283 & 0.356 & 0.347 & 0.338 & \underline{0.399} & \textbf{0.419$^*$} & 17.70\% &\\
\midrule
Home \&  & HR@10   & 0.527 & 0.498 & 0.480  & 0.458 & \underline{0.529} & \textbf{0.565$^*$} & 13.45\%& \multirow{2}{*}{\hphantom{00}9.2} \\
Kitchen  & NDCG@10 & 0.318 & 0.303 & 0.293 & 0.267 & \underline{0.320} & \textbf{0.363$^*$} & 19.80\%& \\
\midrule
\multirow{2}{*}{ML-1M} & HR@10   & 0.440 & \underline{0.819} & 0.804  & 0.802             & 0.608 & \textbf{0.827$^*$} & \hphantom{1}0.97\%& \multirow{2}{*}{292.6} \\
                       & NDCG@10 & 0.241 & 0.594             & 0.576  & \underline{0.603} & 0.351 & \textbf{0.604\hphantom{$^*$}} & \hphantom{1}1.68\%& \\
\midrule
\multirow{2}{*}{Steam} & HR@10  & 0.760 & \underline{0.862} & 0.857 & 0.856             & 0.828 & \textbf{0.871\hphantom{$^*$}} & \hphantom{1}1.04\%& \multirow{2}{*}{322.9} \\
                       & NDCG@10 & 0.500 & 0.639             & 0.635 & \underline{0.645} & 0.610 & \textbf{0.651$^*$} & \hphantom{1}1.88\%& \\
\bottomrule
\end{tabular}
\label{Tab:results}
\end{table}

%% file: figures/cold_start_beauty_fig.tex
\begin{figure*}[ht!]
    \centering
    \hfill
    \begin{subfigure}[b]{.45\textwidth}
			\centering
	   \begin{tikzpicture}[trim axis left,trim axis right]

            \definecolor{color0}{rgb}{0.917647058823529,0.917647058823529,0.949019607843137}
            \definecolor{color1}{rgb}{0.282352941176471,0.470588235294118,0.815686274509804}
            \definecolor{color2}{rgb}{0.933333333333333,0.52156862745098,0.290196078431373}
            \definecolor{color3}{rgb}{0.415686274509804,0.8,0.392156862745098}
            
            \begin{axis}[
            axis line style={black},
            height=5cm,
            width=\columnwidth,
            legend cell align={left},
            legend style={
			},
            legend pos=north west,
            legend entries={{\simrec},{SASRec}},
            tick align=inside,
            tick pos=both,
            grid style={dotted, gray},
		    xlabel={Item Frequency},
            xmajorgrids,
            xmin=0.0, xmax=10,
            xtick style={color=white!15!black},
            xtick={0,2,4,6,8,10},
            xticklabels={
              \(\displaystyle {0}\),
              \(\displaystyle {2}\),
              \(\displaystyle {4}\),
              \(\displaystyle {6}\),
              \(\displaystyle {8}\),
              \(\displaystyle {10+}\)
            },
            ylabel={HR@10},
            ymajorgrids,
            ymin=0, ymax=0.9,
            ytick style={color=white!15!black},
            ytick={0,0.2,0.4,0.6,0.8},
            ylabel near ticks,
            ]

            \addplot [semithick, color3, mark=triangle*, mark options={draw=black}]
            table {%
            0 0.252824
            1 0.307761	
            2 0.369004	
            3 0.397037
            4 0.481343
            5 0.549118		
            6 0.579787	
            7 0.564189	
            8 0.612546
            9 0.629956	
            10 0.881649
            };
            \addplot [semithick, color2, mark=*, mark options={draw=black}]
            table {%
            0 0.0
            1 0.061728
            2 0.155458
            3 0.290419
            4 0.319419
            5 0.390863	
            6 0.496296
            7 0.447284
            8 0.554217
            9 0.602041
            10 0.816149
            };
            \end{axis}
        \end{tikzpicture}
    \end{subfigure}
    \hfill
    \begin{subfigure}[b]{.45\textwidth}
    		\centering
            \begin{tikzpicture}[trim axis left,trim axis right]

            \definecolor{color0}{rgb}{0.917647058823529,0.917647058823529,0.949019607843137}
            \definecolor{color1}{rgb}{0.282352941176471,0.470588235294118,0.815686274509804}
            \definecolor{color2}{rgb}{0.933333333333333,0.52156862745098,0.290196078431373}
            \definecolor{color3}{rgb}{0.415686274509804,0.8,0.392156862745098}
            
            \begin{axis}[
            axis line style={black},
            height=5cm,
            width=\columnwidth,
            legend cell align={left},
            legend style={
			},
            legend pos=north west,
            legend entries={{\simrec},{SASRec}},
            tick align=inside,
            tick pos=both,
            grid style={dotted, gray},
		    xlabel={Item Frequency},
            xmajorgrids,
            xmin=0.0, xmax=10,
            xtick style={color=white!15!black},
            xtick={0,2,4,6,8,10},
            xticklabels={
              \(\displaystyle {0}\),
              \(\displaystyle {2}\),
              \(\displaystyle {4}\),
              \(\displaystyle {6}\),
              \(\displaystyle {8}\),
              \(\displaystyle {10+}\)
            },
            ylabel={NDCG@10},
            ymajorgrids,
            ymin=0, ymax=0.9,
            ytick style={color=white!15!black},
            ytick={0,0.2,0.4,0.6,0.8},
            ylabel near ticks,
            ]

            \addplot [semithick, color3, mark=triangle*, mark options={draw=black}]
            table {%
            0 0.164524
            1 0.202669
            2 0.245000
            3 0.252243	
            4 0.333576	
            5 0.372482			
            6 0.393284	
            7 0.386889 	
            8 0.403298
            9 0.405495	
            10 0.673185
            };
            \addplot [semithick, color2, mark=*, mark options={draw=black}]
            table {%
            0 0.0
            1 0.035311
            2 0.091567	
            3 0.162755	
            4 0.172723
            5 0.237550	
            6 0.319995	
            7 0.271131
            8 0.311396
            9 0.358537
            10 0.576882	 
            };
            \end{axis}
        \end{tikzpicture}
    \end{subfigure}
    \caption{
    Cold-start performance analysis reveals that \simrec{} excels on rare items, showcasing improvements attributed to enhanced performance in such scenarios, particularly evident in the Beauty dataset.
    \simrec{} achieves a remarkable \textbf{78\%} higher HR@10 and a \textbf{101\%} higher NDCG@10 compared to SASRec \cite{kang2018self} for items that appear less than 10 times in the training set (accounting for 61\% of the test set).}
    \label{Fig:cold-start-beauty}
\end{figure*}
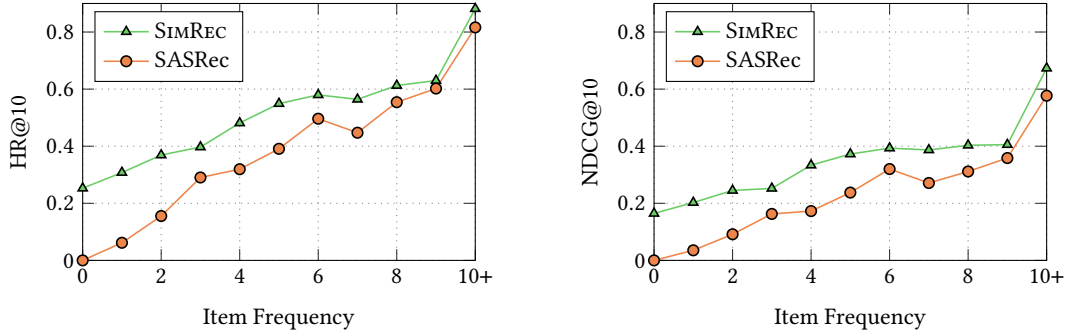

%% file: figures/density_fig.tex
\begin{figure*}[ht!]
    \centering
    \hfill
    \begin{subfigure}[b]{.45\textwidth}
			\centering
	   \begin{tikzpicture}[trim axis left,trim axis right]
            \begin{axis}[
            axis line style={black},
            height=5cm,
            width=\columnwidth,
            tick align=inside,
            tick pos=both,
            grid style={dotted, gray},
		    xlabel={Dataset Density},
            xmajorgrids,
            xmin=3.872268, xmax=15.720102,
            xtick style={color=white!15!black},
            xtick={4,6,8,10,12,14,16},
            ylabel={HR@10 Relative Gain},
            ymajorgrids,
            ymin=0.15, ymax=0.35,
            ytick style={color=white!15!black},
            ytick={0.15,0.2,0.25,0.3,0.35},
            ylabel near ticks,
            yticklabels={
              \(\displaystyle {15\%}\),
              \(\displaystyle {20\%}\),
              \(\displaystyle {25\%}\),
              \(\displaystyle {30\%}\),
              \(\displaystyle {35\%}\),
            },
            ]

            \addplot [semithick, purple!60, mark=*, mark options={draw=black}]
            table {%
            3.872268 0.318372
            6.891064 0.287139
            10.054393 0.231649
            12.963320 0.211946
            15.720102 0.182940
            };
            \end{axis}
        \end{tikzpicture}
    \end{subfigure}
    \hfill
    \begin{subfigure}[b]{.45\textwidth}
    		\centering
	   \begin{tikzpicture}[trim axis left,trim axis right]
            \begin{axis}[
            axis line style={black},
            height=5cm,
            width=\columnwidth,
            tick align=inside,
            tick pos=both,
            grid style={dotted, gray},
		    xlabel={Dataset Density},
            xmajorgrids,
            xmin=3.872268, xmax=15.720102,
            xtick style={color=white!15!black},
            xtick={4,6,8,10,12,14,16},
            ylabel={NDCG@10 Relative Gain},
            ymajorgrids,
            ymin=0.2, ymax=0.45,
            ytick style={color=white!15!black},
            ytick={0.2,0.25,0.3,0.35,0.40,0.45},
            ylabel near ticks,
            yticklabels={
              \(\displaystyle {20\%}\),
              \(\displaystyle {25\%}\),
              \(\displaystyle {30\%}\),
              \(\displaystyle {35\%}\),
              \(\displaystyle {40\%}\),
              \(\displaystyle {45\%}\),
            },
            ]

            \addplot [semithick, purple!60, mark=*, mark options={draw=black}]
            table {%
            3.872268 0.415082
            6.891064 0.365830
            10.054393 0.302551
            12.963320 0.265053
            15.720102 0.226523
            };
            \end{axis}
        \end{tikzpicture}
    \end{subfigure}
    \caption{Relative performance gain across various dataset densities: \simrec{} exhibits a higher relative gain over SASRec \cite{kang2018self}  with the sparser variants of the Beauty dataset, which decreases as dataset density increases.}
    \label{Fig:density}
\end{figure*}
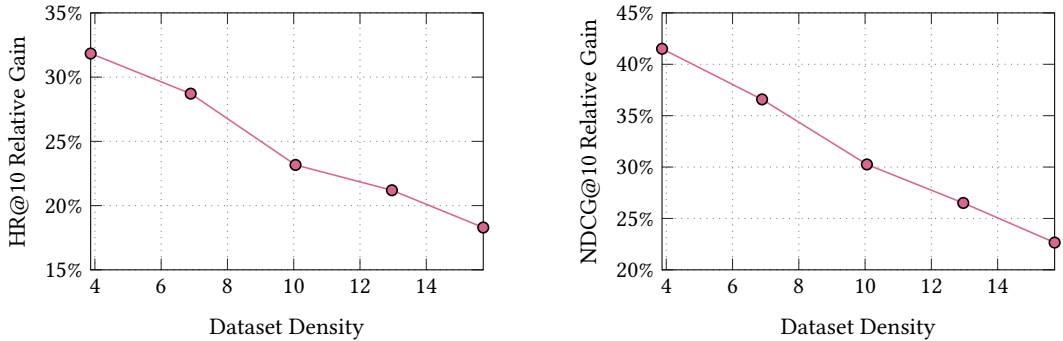

%% file: tables/ablation_table.tex
\begin{table}[h!]
\centering
\caption{Ablation study results. 
}
\vspace{-0.2cm}
\begin{tabular}{lcc}
\toprule

Model & HR@10   & NDCG@10 \\  
\midrule
SASRec  & 0.461 &  0.311 \\
SASRec + embeddings & 0.563  &  0.390 \\
SASRec + embeddings + $\mathcal{L}_{\simrec{}}$ & 0.591  &  0.421 \\
\simrec{} (SASRec + $\mathcal{L}_{\simrec{}}$) & \textbf{0.594} & \textbf{0.425} \\

\bottomrule
\end{tabular}
\label{Tab:ablation}
\end{table}

%% file: 06-concolusion.tex
\section{Conclusions}\label{Se:conclusion}

In this work we present \simrec{}, a new approach to mitigate the cold-start problem in sequential recommendation systems.
\simrec{} exploits the fact that many different items in a dataset are similar to each other, and incorporates similarity information into the training phase using our proposed loss function, $\mathcal{L}_{\simrec{}}$.

\simrec{} outperforms strong sequential recommendation systems like SASRec \cite{kang2018self}, maintaining a similar architecture, thus resulting in the same inference time. Furthermore, it effectively tackles the cold-start problem for sparse datasets, significantly enhancing recommendation performance for items with less than 10 occurrences in the training set, resulting in up to a 78\% improvement in HR@10. 
Notably, \simrec{} was even able to recommend accurately items that were not encountered explicitly during training.

Future research could explore integrating the proposed loss function into diverse sequential recommendation models, alongside investigating alternative similarity functions and diverse data modalities to enhance recommendation performance.

In summary, our approach represents a significant advancement in tackling the challenge of cold-start items with  minimal added efforts and maintaining the same inference time as other sequential recommendation systems.
Notably, it excels in addressing the cold-start problem in sparse datasets, marking a promising direction for future research and applications in recommendation systems.

%% file: appendix.tex
\input{tables/hyperparameter_table}

\section{Implementation Details}\label{Ap:hyper-parameter}

To calculate item embedding for similarity calculation, we used the General Text Embedding Large model (GTE-Large) \cite{li2023towards} -- a pre-trained text embedding model that has been trained with both unsupervised and supervised contrastive objectives to produce dense text representations.
Once we have an embedding vector for each item, we compute the similarities between all items.
Please note that although calculating this process has a quadratic complexity (with the number of items), it occurs only once before training. In practical terms, it takes a relatively short time, less than 1 hour for all six tested datasets, each containing up to 100,000 items, on a typical machine.
This process results in $|\mathcal{V}|^2$ similarity scores, which can be large for some datasets, therefore, we keep only the 1K most similar items for each item.
To reduce noise from less similar items, we also use similarity threshold to filter out items with low similarity.
To maintain a valid similarity distribution over all items $\in \mathcal{V}$, we assign negative infinity to the indices of the filtered out items before applying the softmax function.
We use linear scheduling with warm-up for the hyper parameter $\lambda$ that controls the weight of each component of our loss function.
This approach allows the training process to initially optimize both loss components ($\mathcal{L}_{BCE}$ and $\mathcal{L}_{Similarity}$) and gradually decrease $\lambda$, assigning more weight to $\mathcal{L}_{BCE}$.

We used Adam optimizer \cite{kingma2014adam} with $\beta=ß(0.9, 0.98)$, 210 training epochs, batch size of 128, and 1000 warm-up steps for the linear scheduling of $\lambda$.
We used a maximum sequence length of 200 for ML-1M, and 50 for the other datasets.
The best hyper parameters we found according to the validation set are listed in \Cref{Tab:hyper-parameter}.
When training the other baselines, we used the hyper parameters reported by the authors.

CARCA \cite{rashed2022context} requires additional categorical features. 
For the Beauty, Tools, Pet Supplies, and Home \& Kitchen datasets, we used the price, brand, and categories to which the item belongs. For the ML-1M dataset, we used the year of release and the genres of the movie. For the Steam dataset we used the price, developer and genres.

\input{tables/llms_beauty_results_table}

\section{Text Embedding Method}\label{Se:text-embedding-method}
To find similarities between items, we use a pre-trained text embedding model that produces high-dimensional text embeddings.
To determine the most suitable model, we performed a hyper parameter search (as shown in Table \ref{Tab:hyper-parameter}) on the Beauty dataset using three different models:
\begin{inparaenum}[(a)]
\item GTE-Large \cite{li2023towards},
\item all-MiniLM-L6-v2 \cite{wang2020minilm}, and
\item Sentence-T5-XXL \cite{ni2021sentence}.
\end{inparaenum}
\Cref{Tab:llm-beauty-results} shows the results of the three models.
Since GTE-Large produced the best result, we chose to use it for the other datasets we experimented with.

\input{tables/statistics_table}

\section{Datasets}\label{Se:datasets}
We consider six different datasets --
four of them are part of the Amazon reviews dataset \cite{he2016ups,mcauley2015image}, and the other two consist of movie reviews and video game reviews. 
\begin{enumerate}
    \item \textbf{Tools} \cite{he2016ups,mcauley2015image}: contains items and reviews from the Tools and Home Improvement category.
    \item \textbf{Beauty} \cite{he2016ups,mcauley2015image}: contains items and reviews from the Beauty category.
    \item \textbf{Pet Supplies} \cite{he2016ups,mcauley2015image}: contains items and reviews from the Pet Supplies category.
    \item \textbf{Home and Kitchen} \cite{he2016ups,mcauley2015image}: contains items and reviews from the Home and Kitchen category.
    \item \textbf{ML-1M} \cite{harper2015movielens}: contains 1M reviews from the MovieLens website.
    \item \textbf{Steam} \cite{kang2018self}: contains about 4M reviews from the Steam video games platform. 
\end{enumerate}
Following previous work \cite{kang2018self,rashed2022context,li2020time}, we pre-process the raw datasets and remove items that occur less than $n=5$ times, and then remove users that have less than $5$ interactions.
Note that this pre-processing may lead to items appearing even fewer than $n$ times in the dataset due to the removal of certain users.
We also remove items that do not have a title.
This had a minor effect of removing only 1.3\% of the items on average.
The statistics for the processed datasets can be found in \Cref{Tab:statistics}.

\section{Evaluation Metrics}
To evaluate our approach and other baselines we use the commonly used HR@10 and NDCG@10 metrics.
Following previous work \cite{kang2018self,rashed2022context,li2020time}, for each user interaction history $h_u \in \mathcal{D}$, we consider the first $|u|-2$ items for training, the $|u|-1$ item for validation, and the last item for testing.
Consistent with previous work, we randomly select 100 negative items during the evaluation process given the next true item. Then, we use the model to rank these items and record the HR@10 and NDCG@10 metrics.
Following \citet{rashed2022context}, to reduce the effect of random picking, we repeat the evaluation process 5 times, each time with a set of 100 distinct randomly chosen negative items, and report the average metrics. These evaluation scores are also used to calculate statistical significance compared to the baselines.

\section{Cold-Start Performance}\label{Se:cold-start-additional}
\Cref{Fig:cold-start-beauty-ndcg,Fig:cold-start-tools,Fig:cold-start-pet,Fig:cold-start-homekitchen} show the cold-start improvement obtained by \simrec{} over SASRec \cite{kang2018self}.
On these sparse datasets, \simrec{} exhibits better performance on cold-start items and in particular, items that are not explicitly seen during training.
\Cref{Fig:cold-start-ml1m,Fig:cold-start-steam} show the performance on the dense datasets -- ML-1M and Steam.
For these dense datasets, no significant improvement of one model over the other is evident on cold-start items.
However, in these datasets, less than 1.5\% of the test set samples correspond to rare items that appear less than 10 times in the training set.

\input{figures/cold_start_beauty_with_ndcg_fig}
\input{figures/cold_start_tools_fig}
\input{figures/cold_start_pet_fig}
\input{figures/cold_start_homekitchen_fig}
\input{figures/cold_start_ml1m_fig}
\input{figures/cold_start_steam_fig}

\input{tables/beauty_density_statistics}

\input{tables/density_results_table}

\section{Density vs. Performance -- Additional Information}\label{Se:density-addtional}

The statistics of the dataset variants we generated in \Cref{Se:density-experiments} can be found in \Cref{Tab:beauty-density-statistics}. The exact metric values of the results shown in \Cref{Fig:density} can be found in \Cref{Tab:density_results}.

%% file: tables/hyperparameter_table.tex
\begin{table*}[h]
\centering
\caption{\simrec{} hyper parameters}
\begin{tabular}{llcccccc}
\toprule
\multirow{2}{*}{Parameter} & \multirow{2}{*}{Tested Values} &                    \multicolumn{6}{c}{Best Value}             \\
                            &                                 & Tools & Beauty & Pet Supplies & Home \& Kitchen & ML-1M & Steam\\ 
\midrule
Learning rate        & [1e-4, 2.5e-4, 5e-4, 1e-3] & 1e-4 & 1e-4 & 1e-4 & 1e-4 & 1e-3 &  1e-4\\
Hidden size          & [50, 100]                  & 50   & 100  & 50   & 50   & 100  &  100\\
\# layers            & [2, 3]                     & 3    & 3    & 2    & 3    & 3    &  2\\
Similarity threshold & [0.4, 0.5,..., 1]            & 0.4  & 0.9  & 0.7  & 0.6  & 0.5  &  0.6\\
Softmax temperature          & [0.5, 1, 1.5,..., 4]        & 0.5  & 0.5  & 0.5  & 0.5  & 1    &  1.5\\
$\lambda$            & [0.1, 0.2,..., 0.9]       & 0.8  & 0.3 & 0.6  & 0.7  & 0.6  &  0.2\\
Dropout              & [0.2, 0.5]                 & 0.5  & 0.5  & 0.5  & 0.5  & 0.5  &  0.5\\
\bottomrule
\end{tabular}
\label{Tab:hyper-parameter}
\end{table*}

%% file: tables/llms_beauty_results_table.tex
\begin{table}[h!]
\centering
\caption{Recommendation performance with different text embedding models for item similarity.}
\setlength{\tabcolsep}{5pt}
\begin{tabular}{lcc}
\toprule
Text Embedding Model & HR@10   & NDCG@10 \\
\midrule
Sentence-T5-XXL  & 0.571 &  0.385 \\
all-MiniLM-L6-v2 & 0.581  &  0.410 \\
GTE-Large & \textbf{0.594} & \textbf{0.425} \\
\bottomrule
\end{tabular}
\label{Tab:llm-beauty-results}
\end{table}

%% file: tables/statistics_table.tex
\begin{table}[!ht]
\centering
\caption{Datasets Statistics}
\setlength{\tabcolsep}{5pt}
\begin{tabular}{lrrrr}
\toprule
& Users   & Items & Interactions & Density \\  
\midrule
Tools 	&  45,509  	&  51,186  & 0.31M 	& 6.2\\
Beauty 	&  52,133  	&  57,226  & 0.4M 	& 6.9 \\
Pet Supplies &  33,571  &  27,447  & 0.25M 	&  9.1\\
Home \& Kitchen &  114,724 &   93,358  & 0.86M & 9.2\\
ML-1M &  6,040 &   3,416  & 1M & 292.6\\ 
Steam &  334,634 &   13,045 & 4.2M & 322.9\\ 
\bottomrule
\end{tabular}
\label{Tab:statistics}
\end{table}

%% file: figures/cold_start_beauty_with_ndcg_fig.tex
\begin{figure*}[ht!]
    \centering
    \hfill
    \begin{subfigure}[b]{.45\textwidth}
			\centering
	   \begin{tikzpicture}[trim axis left,trim axis right]

            \definecolor{color0}{rgb}{0.917647058823529,0.917647058823529,0.949019607843137}
            \definecolor{color1}{rgb}{0.282352941176471,0.470588235294118,0.815686274509804}
            \definecolor{color2}{rgb}{0.933333333333333,0.52156862745098,0.290196078431373}
            \definecolor{color3}{rgb}{0.415686274509804,0.8,0.392156862745098}
            
            \begin{axis}[
            axis line style={black},
            height=5cm,
            width=\columnwidth,
            legend cell align={left},
            legend style={
			},
            legend pos=north west,
            legend entries={{\simrec},{SASRec}},
            tick align=inside,
            tick pos=both,
            grid style={dotted, gray},
		    xlabel={Item Frequency},
            xmajorgrids,
            xmin=0.0, xmax=10,
            xtick style={color=white!15!black},
            xtick={0,2,4,6,8,10},
            xticklabels={
              \(\displaystyle {0}\),
              \(\displaystyle {2}\),
              \(\displaystyle {4}\),
              \(\displaystyle {6}\),
              \(\displaystyle {8}\),
              \(\displaystyle {10+}\)
            },
            ylabel={HR@10},
            ymajorgrids,
            ymin=0, ymax=0.9,
            ytick style={color=white!15!black},
            ytick={0,0.2,0.4,0.6,0.8},
            ylabel near ticks,
            ]

            \addplot [semithick, color3, mark=triangle*, mark options={draw=black}]
            table {%
            0 0.252824
            1 0.307761	
            2 0.369004	
            3 0.397037
            4 0.481343
            5 0.549118		
            6 0.579787	
            7 0.564189	
            8 0.612546
            9 0.629956	
            10 0.881649
            };
            \addplot [semithick, color2, mark=*, mark options={draw=black}]
            table {%
            0 0.0
            1 0.061728
            2 0.155458
            3 0.290419
            4 0.319419
            5 0.390863	
            6 0.496296
            7 0.447284
            8 0.554217
            9 0.602041
            10 0.816149
            };
            \end{axis}
        \end{tikzpicture}
    \end{subfigure}
    \hfill
    \begin{subfigure}[b]{.45\textwidth}
    		\centering
            \begin{tikzpicture}[trim axis left,trim axis right]

            \definecolor{color0}{rgb}{0.917647058823529,0.917647058823529,0.949019607843137}
            \definecolor{color1}{rgb}{0.282352941176471,0.470588235294118,0.815686274509804}
            \definecolor{color2}{rgb}{0.933333333333333,0.52156862745098,0.290196078431373}
            \definecolor{color3}{rgb}{0.415686274509804,0.8,0.392156862745098}
            
            \begin{axis}[
            axis line style={black},
            height=5cm,
            width=\columnwidth,
            legend cell align={left},
            legend style={
			},
            legend pos=north west,
            legend entries={{\simrec},{SASRec}},
            tick align=inside,
            tick pos=both,
            grid style={dotted, gray},
		    xlabel={Item Frequency},
            xmajorgrids,
            xmin=0.0, xmax=10,
            xtick style={color=white!15!black},
            xtick={0,2,4,6,8,10},
            xticklabels={
              \(\displaystyle {0}\),
              \(\displaystyle {2}\),
              \(\displaystyle {4}\),
              \(\displaystyle {6}\),
              \(\displaystyle {8}\),
              \(\displaystyle {10+}\)
            },
            ylabel={NDCG@10},
            ymajorgrids,
            ymin=0, ymax=0.9,
            ytick style={color=white!15!black},
            ytick={0,0.2,0.4,0.6,0.8},
            ylabel near ticks,
            ]

            \addplot [semithick, color3, mark=triangle*, mark options={draw=black}]
            table {%
            0 0.164524
            1 0.202669
            2 0.245000
            3 0.252243	
            4 0.333576	
            5 0.372482			
            6 0.393284	
            7 0.386889 	
            8 0.403298
            9 0.405495	
            10 0.673185
            };
            \addplot [semithick, color2, mark=*, mark options={draw=black}]
            table {%
            0 0.0
            1 0.035311
            2 0.091567	
            3 0.162755	
            4 0.172723
            5 0.237550	
            6 0.319995	
            7 0.271131
            8 0.311396
            9 0.358537
            10 0.576882	 
            };
            \end{axis}
        \end{tikzpicture}
    \end{subfigure}
    \caption{Cold-start performance on the Beauty dataset. Datapoints with item frequencies less than 10 collectively account for 61\% of the test set.}
    \label{Fig:cold-start-beauty-ndcg}
\end{figure*}
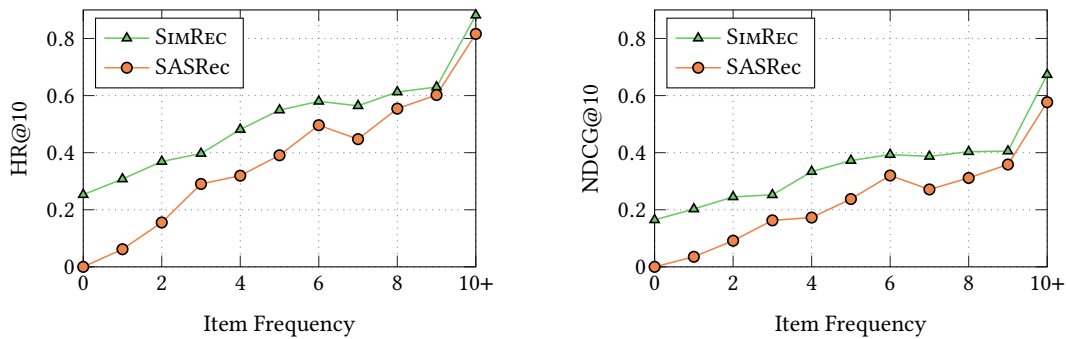

%% file: figures/cold_start_tools_fig.tex
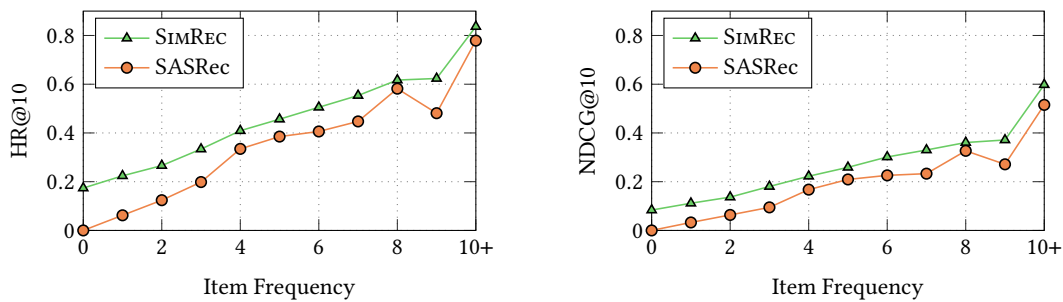
\begin{figure*}[ht!]
    \vspace{0.5cm}
    \centering
    \hfill
    \begin{subfigure}[b]{.45\textwidth}
			\centering
	   \begin{tikzpicture}[trim axis left,trim axis right]

            \definecolor{color0}{rgb}{0.917647058823529,0.917647058823529,0.949019607843137}
            \definecolor{color1}{rgb}{0.282352941176471,0.470588235294118,0.815686274509804}
            \definecolor{color2}{rgb}{0.933333333333333,0.52156862745098,0.290196078431373}
            \definecolor{color3}{rgb}{0.415686274509804,0.8,0.392156862745098}
            
            \begin{axis}[
            axis line style={black},
            height=4.5cm,
            width=\columnwidth,
            legend cell align={left},
            legend style={
			},
            legend pos=north west,
            legend entries={{\simrec},{SASRec}},
            tick align=inside,
            tick pos=both,
            grid style={dotted, gray},
		    xlabel={Item Frequency},
            xmajorgrids,
            xmin=0.0, xmax=10,
            xtick style={color=white!15!black},
            xtick={0,2,4,6,8,10},
            xticklabels={
              \(\displaystyle {0}\),
              \(\displaystyle {2}\),
              \(\displaystyle {4}\),
              \(\displaystyle {6}\),
              \(\displaystyle {8}\),
              \(\displaystyle {10+}\)
            },
            ylabel={HR@10},
            ymajorgrids,
            ymin=0, ymax=0.9,
            ytick style={color=white!15!black},
            ytick={0,0.2,0.4,0.6,0.8},
            ylabel near ticks,
            ]

            \addplot [semithick, color3, mark=triangle*, mark options={draw=black}]
            table {%
            0 0.17384494909945183
            1 0.22407732864674867
            2 0.2661469933184855
            3 0.3338368580060423
            4 0.4091710758377425
            5 0.45647058823529413
            6 0.5053475935828877
            7 0.5536912751677853
            8 0.6170212765957447
            9 0.6235294117647059
            10 0.8364580328814537
            };
            \addplot [semithick, color2, mark=*, mark options={draw=black}]
            table {%
            0 0.0
            1 0.062111801242236024
            2 0.12405609492988134
            3 0.1985185185185185
            4 0.33463796477495106
            5 0.3851203501094092
            6 0.40607734806629836
            7 0.4472843450479233
            8 0.5817490494296578
            9 0.4810606060606061
            10 0.7790165809033733
            };
            \end{axis}
        \end{tikzpicture}
    \end{subfigure}
    \hfill
    \begin{subfigure}[b]{.45\textwidth}
    		\centering
            \begin{tikzpicture}[trim axis left,trim axis right]

            \definecolor{color0}{rgb}{0.917647058823529,0.917647058823529,0.949019607843137}
            \definecolor{color1}{rgb}{0.282352941176471,0.470588235294118,0.815686274509804}
            \definecolor{color2}{rgb}{0.933333333333333,0.52156862745098,0.290196078431373}
            \definecolor{color3}{rgb}{0.415686274509804,0.8,0.392156862745098}
            
            \begin{axis}[
            axis line style={black},
            height=4.5cm,
            width=\columnwidth,
            legend cell align={left},
            legend style={
			},
            legend pos=north west,
            legend entries={{\simrec},{SASRec}},
            tick align=inside,
            tick pos=both,
            grid style={dotted, gray},
		    xlabel={Item Frequency},
            xmajorgrids,
            xmin=0.0, xmax=10,
            xtick style={color=white!15!black},
            xtick={0,2,4,6,8,10},
            xticklabels={
              \(\displaystyle {0}\),
              \(\displaystyle {2}\),
              \(\displaystyle {4}\),
              \(\displaystyle {6}\),
              \(\displaystyle {8}\),
              \(\displaystyle {10+}\)
            },
            ylabel={NDCG@10},
            ymajorgrids,
            ymin=0, ymax=0.9,
            ytick style={color=white!15!black},
            ytick={0,0.2,0.4,0.6,0.8},
            ylabel near ticks,
            ]

            \addplot [semithick, color3, mark=triangle*, mark options={draw=black}]
            table {%
            0 0.08355995841204118
            1 0.11163184156828787
            2 0.13671396768055713
            3 0.18031763745382207
            4 0.22242196083227836
            5 0.25889055726839877
            6 0.3013913967429873
            7 0.33008804340669295
            8 0.3607159131639753
            9 0.37120860923380916
            10 0.5979514506906037
            };
            \addplot [semithick, color2, mark=*, mark options={draw=black}]
            table {%
            0 0.0
            1 0.032712374289879345
            2 0.06359838979991085
            3 0.09459756765526606
            4 0.16745937452979062
            5 0.20921151055112153
            6 0.2259834876461025
            7 0.23298476482578998
            8 0.32670257348499415
            9 0.27127987839439355
            10 0.5152198397817391 
            };
            \end{axis}
        \end{tikzpicture}
    \end{subfigure}
    \vspace{-0.25cm}
    \caption{Cold-start performance on the Tools dataset. Datapoints with item frequencies less than 10 collectively account for 64\% of the test set.}
    \label{Fig:cold-start-tools}
\end{figure*}

%% file: figures/cold_start_pet_fig.tex
\begin{figure*}[ht!]
    \centering
    \hfill
    \begin{subfigure}[b]{.45\textwidth}
			\centering
	   \begin{tikzpicture}[trim axis left,trim axis right]

            \definecolor{color0}{rgb}{0.917647058823529,0.917647058823529,0.949019607843137}
            \definecolor{color1}{rgb}{0.282352941176471,0.470588235294118,0.815686274509804}
            \definecolor{color2}{rgb}{0.933333333333333,0.52156862745098,0.290196078431373}
            \definecolor{color3}{rgb}{0.415686274509804,0.8,0.392156862745098}
            
            \begin{axis}[
            axis line style={black},
            height=4.5cm,
            width=\columnwidth,
            legend cell align={left},
            legend style={
			},
            legend pos=north west,
            legend entries={{\simrec},{SASRec}},
            tick align=inside,
            tick pos=both,
            grid style={dotted, gray},
		    xlabel={Item Frequency},
            xmajorgrids,
            xmin=0.0, xmax=10,
            xtick style={color=white!15!black},
            xtick={0,2,4,6,8,10},
            xticklabels={
              \(\displaystyle {0}\),
              \(\displaystyle {2}\),
              \(\displaystyle {4}\),
              \(\displaystyle {6}\),
              \(\displaystyle {8}\),
              \(\displaystyle {10+}\)
            },
            ylabel={HR@10},
            ymajorgrids,
            ymin=0, ymax=0.9,
            ytick style={color=white!15!black},
            ytick={0,0.2,0.4,0.6,0.8},
            ylabel near ticks,
            ]

            \addplot [semithick, color3, mark=triangle*, mark options={draw=black}]
            table {%
            0 0.1819277108433735
            1 0.2308641975308642
            2 0.3369713506139154
            3 0.41710296684118675
            4 0.4861995753715499
            5 0.49184782608695654
            6 0.5564738292011019
            7 0.6055045871559633
            8 0.6509090909090909
            9 0.6325757575757576
            10 0.8716467189434586
            };
            \addplot [semithick, color2, mark=*, mark options={draw=black}]
            table {%
            0 0.0
            1 0.07195121951219512
            2 0.1657754010695187
            3 0.2692307692307692
            4 0.3826086956521739
            5 0.3925729442970822
            6 0.5068870523415978
            7 0.5229357798165137
            8 0.5475409836065573
            9 0.5708502024291497
            10 0.8282975619920817
            };
            \end{axis}
        \end{tikzpicture}
    \end{subfigure}
    \hfill
    \begin{subfigure}[b]{.45\textwidth}
    		\centering
            \begin{tikzpicture}[trim axis left,trim axis right]

            \definecolor{color0}{rgb}{0.917647058823529,0.917647058823529,0.949019607843137}
            \definecolor{color1}{rgb}{0.282352941176471,0.470588235294118,0.815686274509804}
            \definecolor{color2}{rgb}{0.933333333333333,0.52156862745098,0.290196078431373}
            \definecolor{color3}{rgb}{0.415686274509804,0.8,0.392156862745098}
            
            \begin{axis}[
            axis line style={black},
            height=4.5cm,
            width=\columnwidth,
            legend cell align={left},
            legend style={
			},
            legend pos=north west,
            legend entries={{\simrec},{SASRec}},
            tick align=inside,
            tick pos=both,
            grid style={dotted, gray},
		    xlabel={Item Frequency},
            xmajorgrids,
            xmin=0.0, xmax=10,
            xtick style={color=white!15!black},
            xtick={0,2,4,6,8,10},
            xticklabels={
              \(\displaystyle {0}\),
              \(\displaystyle {2}\),
              \(\displaystyle {4}\),
              \(\displaystyle {6}\),
              \(\displaystyle {8}\),
              \(\displaystyle {10+}\)
            },
            ylabel={NDCG@10},
            ymajorgrids,
            ymin=0, ymax=0.9,
            ytick style={color=white!15!black},
            ytick={0,0.2,0.4,0.6,0.8},
            ylabel near ticks,
            ]

            \addplot [semithick, color3, mark=triangle*, mark options={draw=black}]
            table {%
            0 0.07110315156866787
            1 0.09873098541423313
            2 0.15427850520973338
            3 0.21092831002821566
            4 0.2632824903791845
            5 0.27296382543302966
            6 0.32469332039037496
            7 0.33088424267573796
            8 0.382632873724842
            9 0.39920388107972243
            10 0.6331188233981172
            };
            \addplot [semithick, color2, mark=*, mark options={draw=black}]
            table {%
            0 0.0
            1 0.03389730109824696
            2 0.08609699034960988
            3 0.13982060589605919
            4 0.22038643809629238
            5 0.22696785835626646
            6 0.2945068906000739
            7 0.2999739587564177
            8 0.3156368490016932
            9 0.3516638946494755
            10 0.568323939784101
            };
            \end{axis}
        \end{tikzpicture}
    \end{subfigure}
    \vspace{-0.25cm}
    \caption{Cold-start performance on the Pet Supplies dataset. Datapoints with item frequencies less than 10 collectively account for 51\% of the test set}
    \label{Fig:cold-start-pet}
\end{figure*}

%% file: figures/cold_start_homekitchen_fig.tex
\begin{figure*}[ht!]
    \vspace{0.3cm}
    \centering
    \hfill
    \begin{subfigure}[b]{.45\textwidth}
			\centering
	   \begin{tikzpicture}[trim axis left,trim axis right]

            \definecolor{color0}{rgb}{0.917647058823529,0.917647058823529,0.949019607843137}
            \definecolor{color1}{rgb}{0.282352941176471,0.470588235294118,0.815686274509804}
            \definecolor{color2}{rgb}{0.933333333333333,0.52156862745098,0.290196078431373}
            \definecolor{color3}{rgb}{0.415686274509804,0.8,0.392156862745098}
            
            \begin{axis}[
            axis line style={black},
            height=4.5cm,
            width=\columnwidth,
            legend cell align={left},
            legend style={
			},
            legend pos=north west,
            legend entries={{\simrec},{SASRec}},
            tick align=inside,
            tick pos=both,
            grid style={dotted, gray},
		    xlabel={Item Frequency},
            xmajorgrids,
            xmin=0.0, xmax=10,
            xtick style={color=white!15!black},
            xtick={0,2,4,6,8,10},
            xticklabels={
              \(\displaystyle {0}\),
              \(\displaystyle {2}\),
              \(\displaystyle {4}\),
              \(\displaystyle {6}\),
              \(\displaystyle {8}\),
              \(\displaystyle {10+}\)
            },
            ylabel={HR@10},
            ymajorgrids,
            ymin=0, ymax=0.9,
            ytick style={color=white!15!black},
            ytick={0,0.2,0.4,0.6,0.8},
            ylabel near ticks,
            ]

            \addplot [semithick, color3, mark=triangle*, mark options={draw=black}]
            table {%
0 0.06053268765133172
1 0.11654135338345864
2 0.17061611374407584
3 0.23921568627450981
4 0.3069544364508393
5 0.32122905027932963
6 0.43788819875776397
7 0.48148148148148145
8 0.5301724137931034
9 0.572139303482587
10 0.8349885408708938
            };
            \addplot [semithick, color2, mark=*, mark options={draw=black}]
            table {%
0 0.0
1 0.026615969581749048
2 0.06334841628959276
3 0.13562753036437247
4 0.19811320754716982
5 0.22070844686648503
6 0.32840236686390534
7 0.375
8 0.3902439024390244
9 0.4031413612565445
10 0.8018104776579353
            };
            \end{axis}
        \end{tikzpicture}
    \end{subfigure}
    \hfill
    \begin{subfigure}[b]{.45\textwidth}
    		\centering
            \begin{tikzpicture}[trim axis left,trim axis right]

            \definecolor{color0}{rgb}{0.917647058823529,0.917647058823529,0.949019607843137}
            \definecolor{color1}{rgb}{0.282352941176471,0.470588235294118,0.815686274509804}
            \definecolor{color2}{rgb}{0.933333333333333,0.52156862745098,0.290196078431373}
            \definecolor{color3}{rgb}{0.415686274509804,0.8,0.392156862745098}
            
            \begin{axis}[
            axis line style={black},
            height=4.5cm,
            width=\columnwidth,
            legend cell align={left},
            legend style={
			},
            legend pos=north west,
            legend entries={{\simrec},{SASRec}},
            tick align=inside,
            tick pos=both,
            grid style={dotted, gray},
		    xlabel={Item Frequency},
            xmajorgrids,
            xmin=0.0, xmax=10,
            xtick style={color=white!15!black},
            xtick={0,2,4,6,8,10},
            xticklabels={
              \(\displaystyle {0}\),
              \(\displaystyle {2}\),
              \(\displaystyle {4}\),
              \(\displaystyle {6}\),
              \(\displaystyle {8}\),
              \(\displaystyle {10+}\)
            },
            ylabel={NDCG@10},
            ymajorgrids,
            ymin=0, ymax=0.9,
            ytick style={color=white!15!black},
            ytick={0,0.2,0.4,0.6,0.8},
            ylabel near ticks,
            ]

            \addplot [semithick, color3, mark=triangle*, mark options={draw=black}]
            table {%
0 0.027997027513947708
1 0.05942755520286587
2 0.08966133589928174
3 0.1224408135280541
4 0.15823998421246266
5 0.1738130128813852
6 0.24501733485638524
7 0.2626313272373209
8 0.3166614697860584
9 0.3338369979345198
10 0.5647314486029902
            };
            \addplot [semithick, color2, mark=*, mark options={draw=black}]
            table {%
0 0.0
1 0.012056491491877
2 0.03285634232325958
3 0.07072560568627514
4 0.09550109105021434
5 0.11613737255386469
6 0.16996561353360265
7 0.16673238310531904
8 0.19955990103545532
9 0.22346658199745298
10 0.5007414602227286
            };
            \end{axis}
        \end{tikzpicture}
    \end{subfigure}
    \vspace{-0.25cm}
    \caption{Cold-start performance on the Home \& Kitchen dataset. Datapoints with item frequencies less than 10 collectively account for 47\% of the test set.}
    \label{Fig:cold-start-homekitchen}
\end{figure*}

%% file: figures/cold_start_ml1m_fig.tex
\begin{figure*}[ht!]
    \vspace{0.3cm}
    \centering
    \hfill
    \begin{subfigure}[b]{.45\textwidth}
			\centering
	   \begin{tikzpicture}[trim axis left,trim axis right]

            \definecolor{color0}{rgb}{0.917647058823529,0.917647058823529,0.949019607843137}
            \definecolor{color1}{rgb}{0.282352941176471,0.470588235294118,0.815686274509804}
            \definecolor{color2}{rgb}{0.933333333333333,0.52156862745098,0.290196078431373}
            \definecolor{color3}{rgb}{0.415686274509804,0.8,0.392156862745098}
            
            \begin{axis}[
            axis line style={black},
            height=4.5cm,
            width=\columnwidth,
            legend cell align={left},
            legend style={
			},
            legend pos=north west,
            legend entries={{SASRec}, {\simrec}},
            reverse legend,
            tick align=inside,
            tick pos=both,
            grid style={dotted, gray},
		    xlabel={Item Frequency},
            xmajorgrids,
            xmin=4.0, xmax=10,
            xtick style={color=white!15!black},
            xtick={4,6,8,10},
            xticklabels={
              \(\displaystyle {4}\),
              \(\displaystyle {6}\),
              \(\displaystyle {8}\),
              \(\displaystyle {10+}\)
            },
            ylabel={HR@10},
            ymajorgrids,
            ymin=0, ymax=0.9,
            ytick style={color=white!15!black},
            ytick={0,0.2,0.4,0.6,0.8},
            ylabel near ticks,
            ]

            \addplot [semithick, color2, mark=*, mark options={draw=black}]
            table {%
4 0.0
5 0.0
6 0.16666666666666666
7 0.0
8 0.0
9 0.3333333333333333
10 0.824447766151802
            };
            \addplot [semithick, color3, mark=triangle*, mark options={draw=black}]
            table {%
4 0.0
5 0.0
6 0.16666666666666666
7 0.0
8 0.0
9 0.3333333333333333
10 0.8249460222554393
            };
            \end{axis}
        \end{tikzpicture}
    \end{subfigure}
    \hfill
    \begin{subfigure}[b]{.45\textwidth}
    		\centering
            \begin{tikzpicture}[trim axis left,trim axis right]

            \definecolor{color0}{rgb}{0.917647058823529,0.917647058823529,0.949019607843137}
            \definecolor{color1}{rgb}{0.282352941176471,0.470588235294118,0.815686274509804}
            \definecolor{color2}{rgb}{0.933333333333333,0.52156862745098,0.290196078431373}
            \definecolor{color3}{rgb}{0.415686274509804,0.8,0.392156862745098}
            
            \begin{axis}[
            axis line style={black},
            height=4.5cm,
            width=\columnwidth,
            legend cell align={left},
            legend style={
			},
            legend pos=north west,
            legend entries={{SASRec}, {\simrec}},
            reverse legend,
            tick align=inside,
            tick pos=both,
            grid style={dotted, gray},
		    xlabel={Item Frequency},
            xmajorgrids,
            xmin=4.0, xmax=10,
            xtick style={color=white!15!black},
            xtick={4,6,8,10},
            xticklabels={
              \(\displaystyle {4}\),
              \(\displaystyle {6}\),
              \(\displaystyle {8}\),
              \(\displaystyle {10+}\)
            },
            ylabel={NDCG@10},
            ymajorgrids,
            ymin=0, ymax=0.9,
            ytick style={color=white!15!black},
            ytick={0,0.2,0.4,0.6,0.8},
            ylabel near ticks,
            ]

            \addplot [semithick, color2, mark=*, mark options={draw=black}]
            table {%
4 0.0
5 0.0
6 0.050171665943996864
7 0.0
8 0.0
9 0.16666666666666666
10 0.5990739402564446
            };
            \addplot [semithick, color3, mark=triangle*, mark options={draw=black}]
            table {%
4 0.0
5 0.0
6 0.05555555555555555
7 0.0
8 0.0
9 0.1289509357448472
10 0.604484972205901
            };

            \end{axis}
        \end{tikzpicture}
    \end{subfigure}
    \vspace{-0.25cm}
    \caption{Cold-start performance on the ML-1M dataset. Datapoints with item frequencies less than 10 collectively account for 0.3\% of the test set.}
    \label{Fig:cold-start-ml1m}
\end{figure*}

%% file: figures/cold_start_steam_fig.tex
\begin{figure*}[ht!]
    \vspace{0.3cm}
    \centering
    \hfill
    \begin{subfigure}[b]{.45\textwidth}
			\centering
	   \begin{tikzpicture}[trim axis left,trim axis right]

            \definecolor{color0}{rgb}{0.917647058823529,0.917647058823529,0.949019607843137}
            \definecolor{color1}{rgb}{0.282352941176471,0.470588235294118,0.815686274509804}
            \definecolor{color2}{rgb}{0.933333333333333,0.52156862745098,0.290196078431373}
            \definecolor{color3}{rgb}{0.415686274509804,0.8,0.392156862745098}
            
            \begin{axis}[
            axis line style={black},
            height=4.5cm,
            width=\columnwidth,
            legend cell align={left},
            legend style={
			},
            legend pos=north west,
            legend entries={{SASRec}, {\simrec}},
            reverse legend,
            tick align=inside,
            tick pos=both,
            grid style={dotted, gray},
		    xlabel={Item Frequency},
            xmajorgrids,
            xmin=0.0, xmax=10,
            xtick style={color=white!15!black},
            xtick={0,2,4,6,8,10},
            xticklabels={
              \(\displaystyle {0}\),              
              \(\displaystyle {2}\),
              \(\displaystyle {4}\),
              \(\displaystyle {6}\),
              \(\displaystyle {8}\),
              \(\displaystyle {10+}\)
            },
            ylabel={HR@10},
            ymajorgrids,
            ymin=0, ymax=0.9,
            ytick style={color=white!15!black},
            ytick={0,0.2,0.4,0.6,0.8},
            ylabel near ticks,
            ]

            \addplot [semithick, color2, mark=*, mark options={draw=black}]
            table {%
0 0.0
1 0.0
2 0.0
3 0.0
4 0.11764705882352941
5 0.058823529411764705
6 0.07142857142857142
7 0.058823529411764705
8 0.25
9 0.21428571428571427
10 0.8773833671399595
            };
            \addplot [semithick, color3, mark=triangle*, mark options={draw=black}]
            table {%
0 0.0
1 0.0
2 0.0
3 0.0
4 0.043478260869565216
5 0.1
6 0.07142857142857142
7 0.047619047619047616
8 0.2
9 0.1875
10 0.8817794028031688
            };
            \end{axis}
        \end{tikzpicture}
    \end{subfigure}
    \hfill
    \begin{subfigure}[b]{.45\textwidth}
    		\centering
            \begin{tikzpicture}[trim axis left,trim axis right]

            \definecolor{color0}{rgb}{0.917647058823529,0.917647058823529,0.949019607843137}
            \definecolor{color1}{rgb}{0.282352941176471,0.470588235294118,0.815686274509804}
            \definecolor{color2}{rgb}{0.933333333333333,0.52156862745098,0.290196078431373}
            \definecolor{color3}{rgb}{0.415686274509804,0.8,0.392156862745098}
            
            \begin{axis}[
            axis line style={black},
            height=4.5cm,
            width=\columnwidth,
            legend cell align={left},
            legend style={
			},
            legend pos=north west,
            legend entries={{SASRec}, {\simrec}},
            reverse legend,
            tick align=inside,
            tick pos=both,
            grid style={dotted, gray},
		    xlabel={Item Frequency},
            xmajorgrids,
            xmin=0.0, xmax=10,
            xtick style={color=white!15!black},
            xtick={0,2,4,6,8,10},
            xticklabels={
              \(\displaystyle {0}\),              
              \(\displaystyle {2}\),
              \(\displaystyle {4}\),
              \(\displaystyle {6}\),
              \(\displaystyle {8}\),
              \(\displaystyle {10+}\)
            },
            ylabel={NDCG@10},
            ymajorgrids,
            ymin=0, ymax=0.9,
            ytick style={color=white!15!black},
            ytick={0,0.2,0.4,0.6,0.8},
            ylabel near ticks,
            ]

            \addplot [semithick, color2, mark=*, mark options={draw=black}]
            table {%
0 0.0
1 0.0
2 0.0
3 0.0
4 0.034711460116580536
5 0.025333915180787828
6 0.030762611290956646
7 0.0177076468037636
8 0.0934822778215063
9 0.1208954332297363
10 0.6519947333632291
            };
            \addplot [semithick, color3, mark=triangle*, mark options={draw=black}]
            table {%
0 0.0
1 0.0
2 0.0
3 0.0
4 0.01308826068104266
5 0.04087646826539674
6 0.020647487594134848
7 0.018421562249263886
8 0.08175293653079348
9 0.06635039447780317
10 0.6607317339060641
            };

            \end{axis}
        \end{tikzpicture}
    \end{subfigure}
    \vspace{-0.25cm}
    \caption{Cold-start performance on the Steam dataset. Datapoints with item frequencies less than 10 collectively account for 1.5\% of the test set.}
    \label{Fig:cold-start-steam}
\end{figure*}
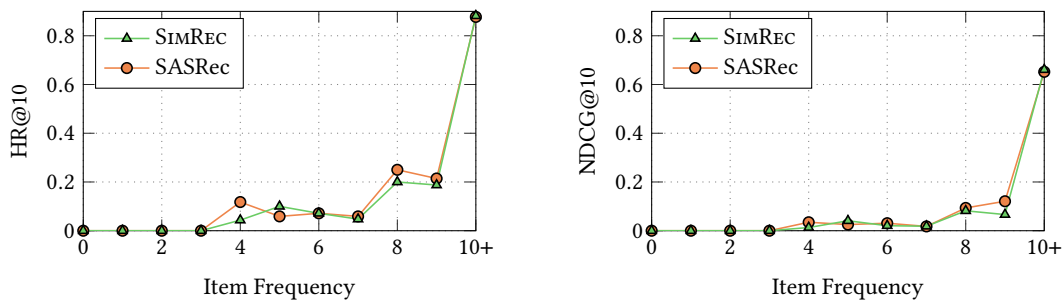

%% file: tables/beauty_density_statistics.tex
\begin{table*}[!ht]
\centering
\caption{Variants of the Beauty dataset with different densities. Items appearing less than $n$ times are excluded from the dataset. $n=5$ corresponds to the dataset we present in \Cref{Se:datasets}.}
\setlength{\tabcolsep}{5pt}
\begin{tabular}{lrrrr}
\toprule
$n$ & Users   & Items & Interactions & Density \\  
\midrule
0  &  52,302 &  121,144 & 0.47M & 3.9\\
5  &  52,133 &  57,226  & 0.4M  & 6.9 \\
10 &  51,817 &  34,784  & 0.35M & 10.1 \\
15 &  51,433 &  24,618  & 0.32M & 13.0\\
20 &  51,004  & 18,764  & 0.3M  & 15.7\\ 
\bottomrule
\end{tabular}
\label{Tab:beauty-density-statistics}
\end{table*}

%% file: tables/density_results_table.tex
\begin{table*}[h!]
\centering
\caption{Recommendation performance on variable density levels of the Beauty dataset. Items that appear less than $n$ times in the raw dataset are discarded.}
\begin{tabular}{llcccc}
\toprule
$n$ & Metric & SASRec & \simrec{} & Improve vs. SASRec & Density \\
\midrule
\multirow{2}{*}{0} & HR@10   & 0.455 & 0.599 & 31.84\% & \multirow{2}{*}{3.9}\\
                   & NDCG@10 & 0.310 & 0.439 & 41.51\% &\\
\midrule                 
\multirow{2}{*}{5} & HR@10   & 0.461 & 0.594 & 28.85\% & \multirow{2}{*}{6.9}\\
                   & NDCG@10 & 0.311 & 0.425 & 36.70\% &\\
\midrule
\multirow{2}{*}{10} & HR@10   & 0.463 & 0.571 & 23.16\% & \multirow{2}{*}{10.1}\\
                    & NDCG@10 & 0.301 & 0.403 & 30.26\% &\\
\midrule
\multirow{2}{*}{15} & HR@10   & 0.467 & 0.566 & 21.19\% & \multirow{2}{*}{13.0}\\
                    & NDCG@10 & 0.314 & 0.397 & 26.51\% &\\
\midrule
\multirow{2}{*}{20} & HR@10   & 0.470 & 0.556 & 18.29\% & \multirow{2}{*}{15.7}\\
                    & NDCG@10 & 0.315 & 0.387 & 22.65\% &\\
\bottomrule
\end{tabular}
\label{Tab:density_results}
\end{table*}